# Prominence Cavity Regions Observed using SWAP 174Å Filtergrams and Simultaneous Eclipse Flash Spectra

by:

C. Bazin . S. Koutchmy . E. Tavabi*

Institut d'Astrophysique de Paris, UMR 7095, CNRS and UPMC Paris, France

* Physics Department, Payame Noor University, 19395-3697 Teheran  Iran

email: bazin@iap.fr

**Abstract**:  SWAP images from PROBA2 taken at 174 Å in the Fe IX/X lines are compared with simultaneous slitless flash spectra obtained during the solar total eclipse of 11 July, 2010.  Myriads of faint low-excitation emission lines together with the He I and He II Paschen α chromospheric lines are recorded on eclipse spectra where regions of limb prominences are obtained with space-borne imagers. We analyzed a deep flash spectrum obtained by summing 80 individual spectra to evaluate the intensity modulations of the continuum. Intensity deficits are observed and measured at the prominences boundaries in both eclipse and SWAP images. The prominence cavities interpreted as a relative depression of plasma density, produced inside the corona surrounding the prominences, and some intense heating occurring in these regions, are discussed. Photometric measurements are shown at different scales and different, spectrally narrow, intervals for both the prominences and the coronal background.

**Key words**: Prominences; Prominence- Corona Interface; Helium spectra

## 1) Introduction

 Prominences and their disk counterpart, filaments, rise to coronal height and are cool objects formed above a polarity inversion line (PIL) of the large-scale photospheric magnetic field. Their thermodynamic properties are well described by a standard model (Engvold *et al.* 1990) showing a large dispersion of the critical parameters; the latest developments of both high-resolution groundbased and space observations still show several open issues (Labrosse *et al.* 2010). White-light (WL) photographic eclipse images have shown features around prominences that have been called "cavity separations" when observed in projection on the plane of the sky. The prominence and the overlying system of arches with the streamer sheet extend radially to several radii (Saito and Hyder, 1968, Saito and Tandberg-Hanssen 1973, Tandberg-Hanssen, 1995). Sometimes the filament/prominence/cavity system erupts, so that the analysis of these regions, including the origin of the heating presumably responsible for the eruption or accompanying it, is of great interest. Broad-band WL eclipse CCD or CMOS images, taken in different wavelengths, now also show the limb prominences (November and Koutchmy, 1996; Koutchmy, Filippov and Lamy, 2007; Jejcic and Heinzel, 2009), allowing a more quantitative analysis. The use of eclipse slitless spectra, taken with a modern detector, makes the spectral analysis possible, including the measurement of properly filtered WL radiation. Extreme ultra-violet (EUV) filtergrams, obtained in space well before and after the eclipse, permit a more extended temporal analysis of these regions, including the dynamical effects and the 3D geometry, taking advantage of the change in viewing angle, due to solar rotation. This allows the description of the behavior of the corona around prominences and the structure of the magnetic field in the context of the filament channel, the PIL, and the streamer association. However, it is clear that the lack of spatio -temporal resolution at the scale of interest for the heating (of order of 100 km or less and a few seconds of time or less) is a limitation that should be kept in mind when chronologic dynamical phenomena are discussed, in addition to the lack of relevant direct magnetic-field measurements in these coronal regions.

The total eclipse of 11 July 2010 (see Pasachoff *et al.* 2011; Habbal *et al.* 2010, 2011 for a description of the corona at this eclipse), allowed us to observe the true continuum between helium prominence lines (He I 4713 Å and Pα He II 4686 Å), taking into account faint low-excitation emission lines using fast slitless spectra taken at the contacts, before and after the totality. A sketch of the experiment is given in the appendix (Figure 15) (from Bazin *et al* 2011). These observations also yield relatively new results on the determination of the electron densities inside prominences when compared to previous photographic ground-based coronagraphic filtergram observations in the $D_3$ line, where the continuum, at the location of the prominences, was not precisely measured (Kubota and Leroy, 1970 for Lyot-coronagraph observations and Koutchmy, Lebecq, and Stellmacher 1983 for eclipse broadband observations). Flash spectra, measuring the continuum outside prominences, allow one to study the electron density of the cavity and compare with the PROBA2/ Sun Watcher with APS detectors and image Processing (SWAP*)* EUV filtergrams, dominantly recording the Fe IX/X coronal lines (Seaton *et al* 2012, SWAP instrument paper, this issue). SWAP is one of the main scientific instruments of the mission Project for OnBoard Autonomy (PROBA2: Berghmans *et al*. 2006; Defise *et al*. 2007; Halain *et al.* 2010, De Groof *et al*, 2008a, 2008b for the CMOS-APS imaging detectors). PROBA2 is an ESA microsatellite launched in 2009. The SWAP imager has a field of view of 54x54 arc minutes which is substantially wider than the field of view of the well-known EIT imager of SOHO, and has a typical cadence of one image per minute.

The presence of many faint low excitation emission lines in our flash spectra[1], such as lines of Ti +, Fe +, and Mg+, could eventually reveal the importance of the first ionization potential

---

[1] From the historical point of view, flash spectra were rather long exposure frames using photographic plates at the exit of a slitless spectrograph. They were taken just after the last Baily's beads seen in white light,

(FIP) effect which occurs in the low layers of the transition region (TR), but their discussion is beyond the scope of this article. Note that these faint emission lines, immediately outside the limb of the Sun, are impossible to observe properly outside total solar eclipses, due to the effect of parasitic scattered light of instrumental origin, even with a Lyot coronagraph which usually over-occults the Sun.

A mapping of the intensity ratio of He I/He II lines (at 4713 Å and 4686 Å) inside the SWAP cavity images will also be used for the study of its distribution, as well as for the temperature evaluation of the corona-prominence region, where rather surprising coronagraphic temperature measurements were reported in the past, based on these lines and their profiles (Hirayama and Nakagomi, 1974; Hirayama and Irie, 1984). Here we have the advantage of having many WL images of the corona taken during the eclipse totality (Pasachoff *et al* 2011; Habbal *et al*., 2011).

**2) Observations.**

**2.1) Global View of the Corona during the 11 July 2011 Total Eclipse.**

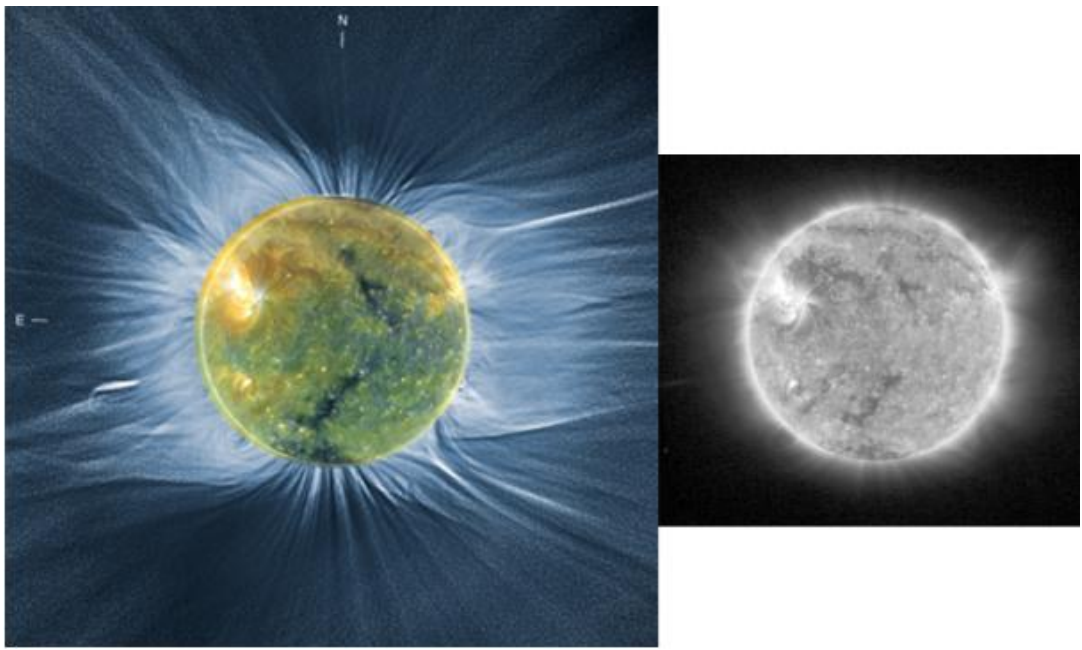

1-a                                1-b

Figure 1a: Composite of a processed WL total eclipse image of the corona on which is superposed a composite AIA image in 171, 193 and 211 Å made almost simultaneously,

---

when the chromospheric lines are still faint and are superposed on the myriad narrow low excitation emission lines overlying the very faint photospheric continuum left outside the solar limb. The faintest lines are revealed at the coronal intensity level during total eclipse, when the level of parasitic scattered light is completely negligible.

with proper scaling and orientation. Figure 1b: Quasi- simultaneous SWAP stacked filtergram at 174 Å, corresponding to the totality of the 11 July 2011 eclipse observed from Hao (French Polynesia), at the same scale.

The ground based experiments carried out in French Polynesia produced several images taken in WL. They allow identification of the streamers and arches (see FIGURE 1a) with an image taken at 18:43 UT and a SWAP stacked EUV filtergram taken near 19:00 UT (Figure 1b). From Figure 1a and 1b, arches above limb prominences were identified, especially the one at the southeast limb, where a cavity is seen in WL with the corresponding prominence inside, which also is observed in the helium lines using our flash spectra. The cavities are dark and surround the prominences close to the limb. More data were obtained thanks to the high-cadence flash spectra, which allow a precise evaluation of the heights above the lunar profile. The emission lines could then be measured from the resulting light flux, seen above the edge of the Moon, at the low heights where the magnetic field is known to dominate the kinetic pressure: the plasma β is lower than one.

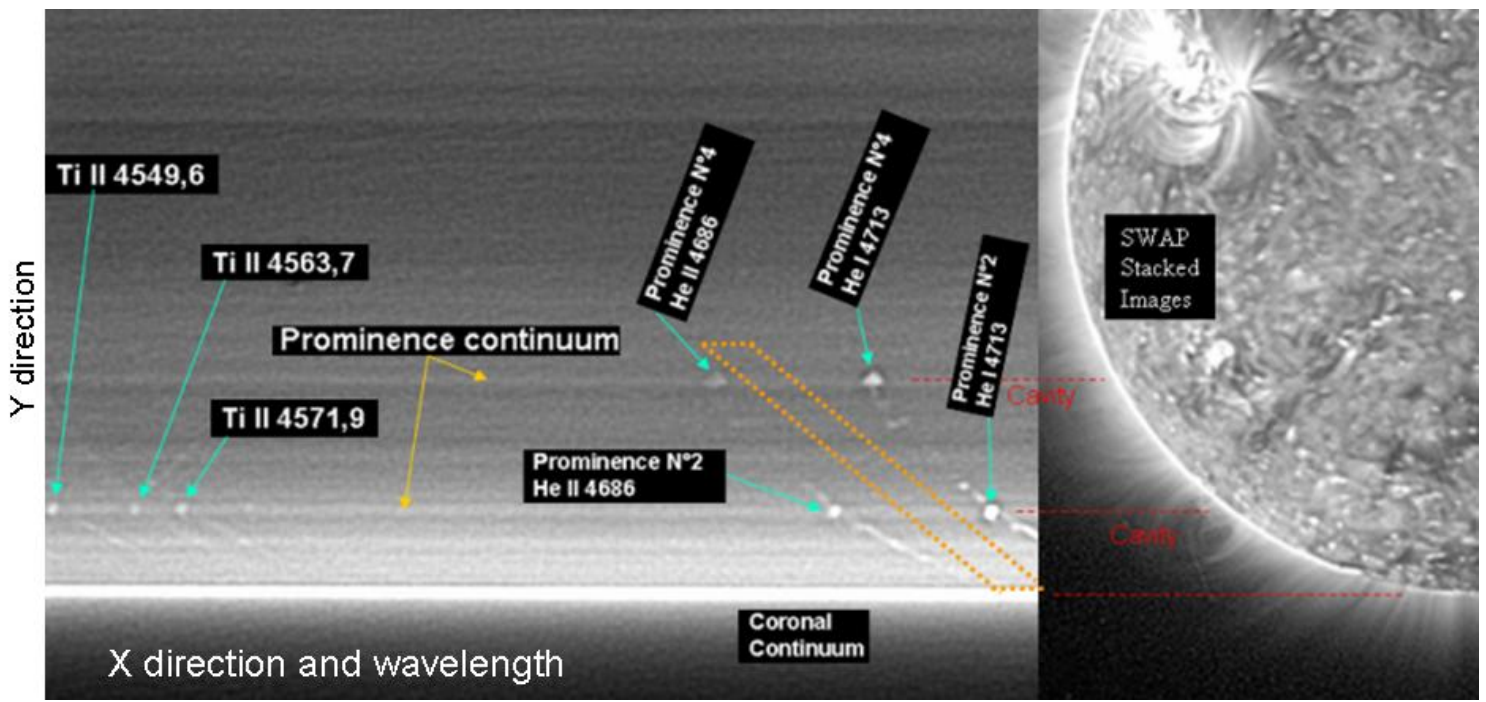

Figure 2: Processed partial image from the flash-spectra sequence, showing the continuum between helium-line prominence emissions. An eight-seconds integration was used (from images taken near the second contact at Hao, from 18:41:53 to 18:42:01), to show the cavity effect (decreased fluxes). The enhancement due to the prominences seen in the true continuum (at left) is compared to a partial SWAP processed image, taken in the same location at the southeast limb (at right). The orientation and scale of the SWAP image are the same as in the spectrum. The dashed-orange lines show the region used to extract the photometric profiles of the continuum which is plotted on Figures 4a and 4b. It is situated

between the He I and He II emission lines. Note that wavelengths considerably drift when changing of Y positions because the equivalent curved slit follows the limb where intensities largely dominate and the dispersion is linear.

We also performed photometric cuts outside the helium emissions, in the continuum, along the direction shown with dashed orange lines in Figure 2, to deduce the intensity variations in the prominence-corona region. Figure 2 shows the comparison of a composite made of 80 stacked eclipse flash-spectra with the SWAP filtergram taken at the time of the eclipse totality; the location where the photometric cuts were performed is indicated. The location of the prominences in the surrounding corona are given in Figure 3.

**2.2) Analysis of the Prominence-Corona region using SWAP, AIA and Flash Spectra in the Region of the 4686 Å He II line.**

The cavity looks relatively dark in WL due to a reduced electron or plasma density (Saito and Tandberg-Hanssen, 1973). A simple interpretation of this long duration phenomenon would be to consider that the missing mass fills the prominence magnetic structure. This question is still open and linked to the so-called prominence-corona interface (PCI) physics, to small-scale plasma flows around the prominence and to the topology of the magnetic field (Hirayama, 1964). It could be the result of an intermittent condensation process of suddenly, radiatively cooled coronal gas, falling towards the complex prominence magnetic structure imbeded into the surrounding corona. This process occurs at very small scales (Koutchmy, Filippov, and Lamy, 2007), where turbulent motions are usually assumed. New eclipse observations should now be compared with space-borne EUV filtergrams obtained simultaneously, in order to further investigate the former idea. The following images were obtained by the *Atmospheric Imaging Assembly* AIA of the *Solar Dynamic Observatory* (SDO). SDO/AIA takes one image every 12 seconds in each wavelength of coronal lines (see http://sdo.gsfc.nasa.gov/) with a considerably improved spatial resolution but a field of view somewhat smaller than that of SWAP. Prominences are seen in emission in the He II 304 Å resonance line on AIA filtergrams and, simultaneously, in the ionized helium Pα line of He II at 4686 Å, using our flash spectra obtained during the total eclipse. Note that the Pα 4686 Å line is considerably less optically thick than the 304 Å resonance line. (Labrosse *et al*, 2010).

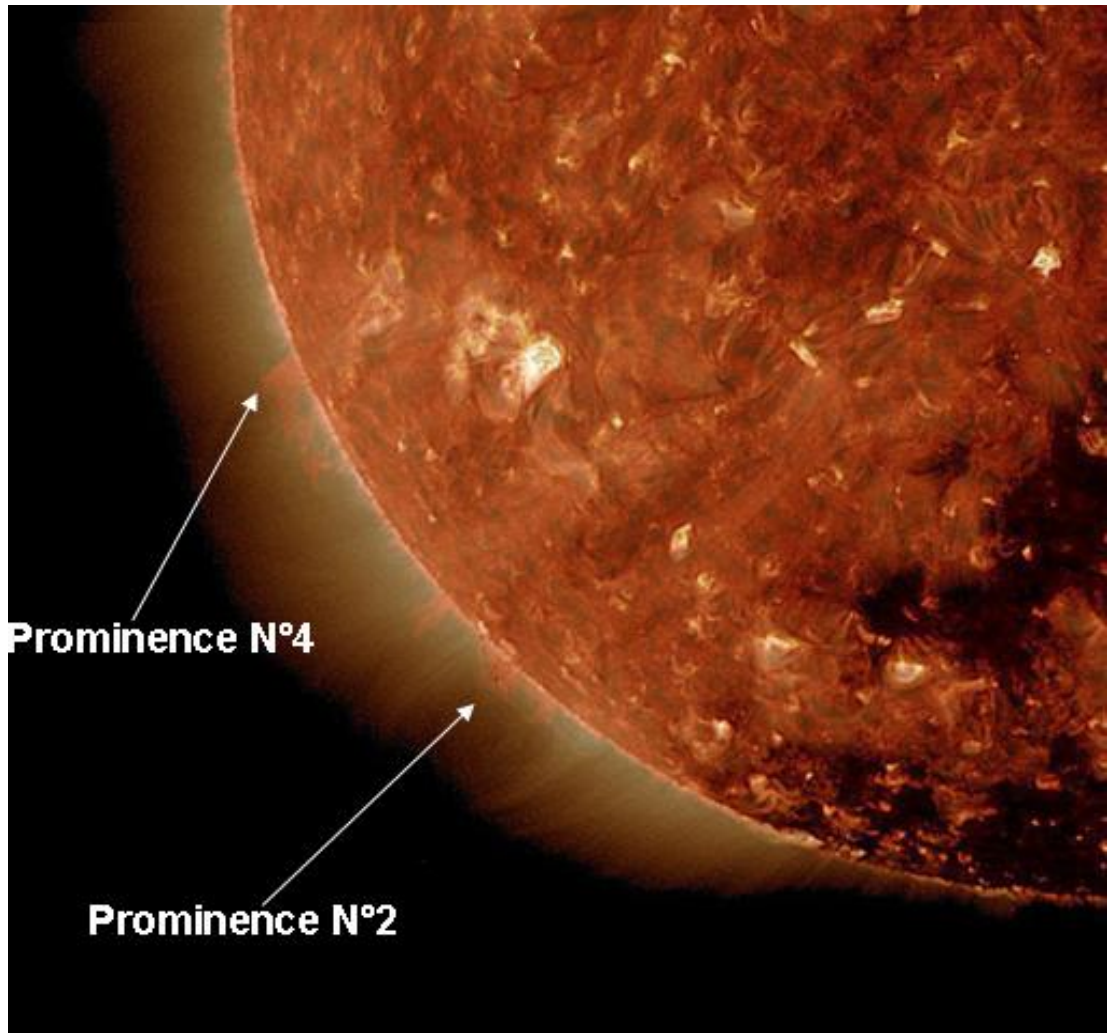

Figure 3: Superposition of the 304 Å and the 193 Å emissions in the partial frame image at southeast taken at the time of the eclipse using SDO/AIA. The arrows point to the prominences studied.

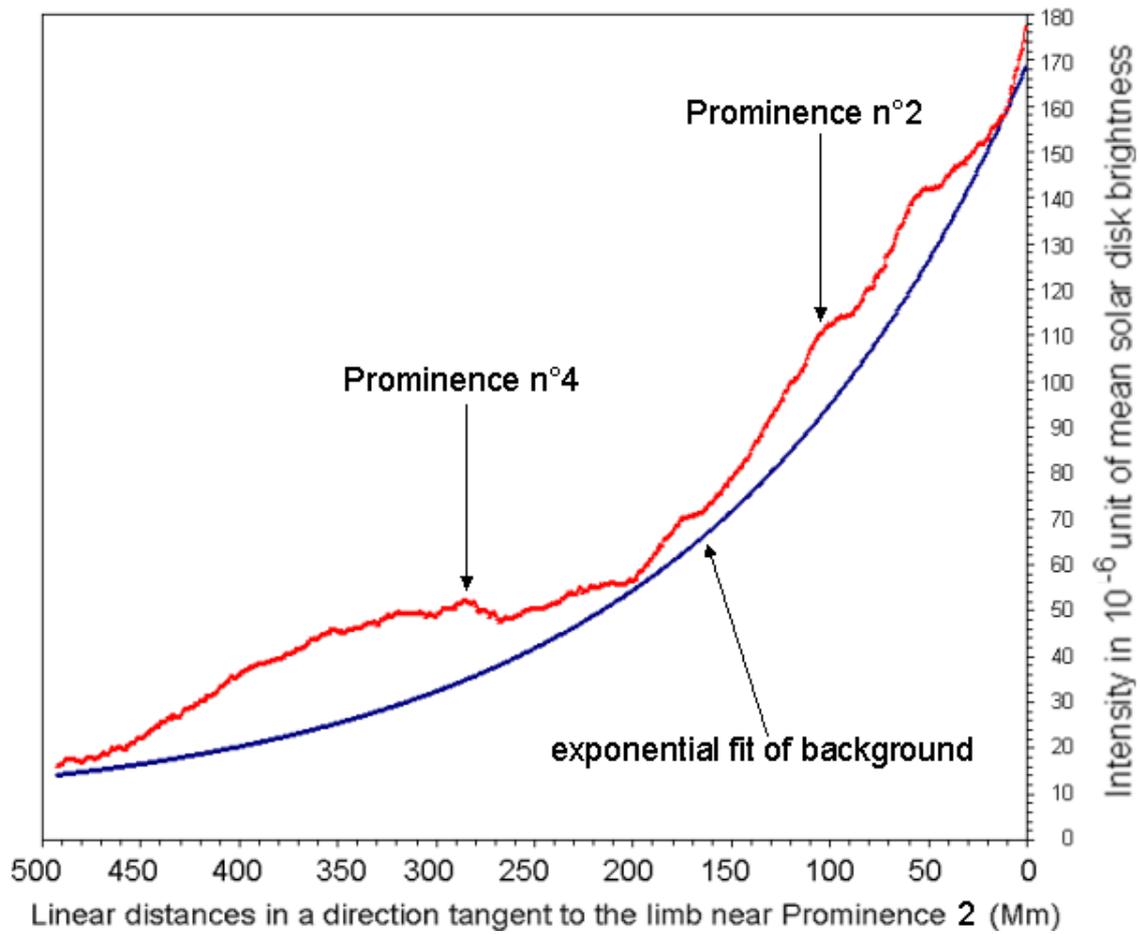

Figure 4a: Photometric cut through the flash spectrum corresponding to the WL "clean" continuum dashed region shown on Figure 2. The blue curve represents the assumed homogeneous background that we used to deduce Figure 4b.

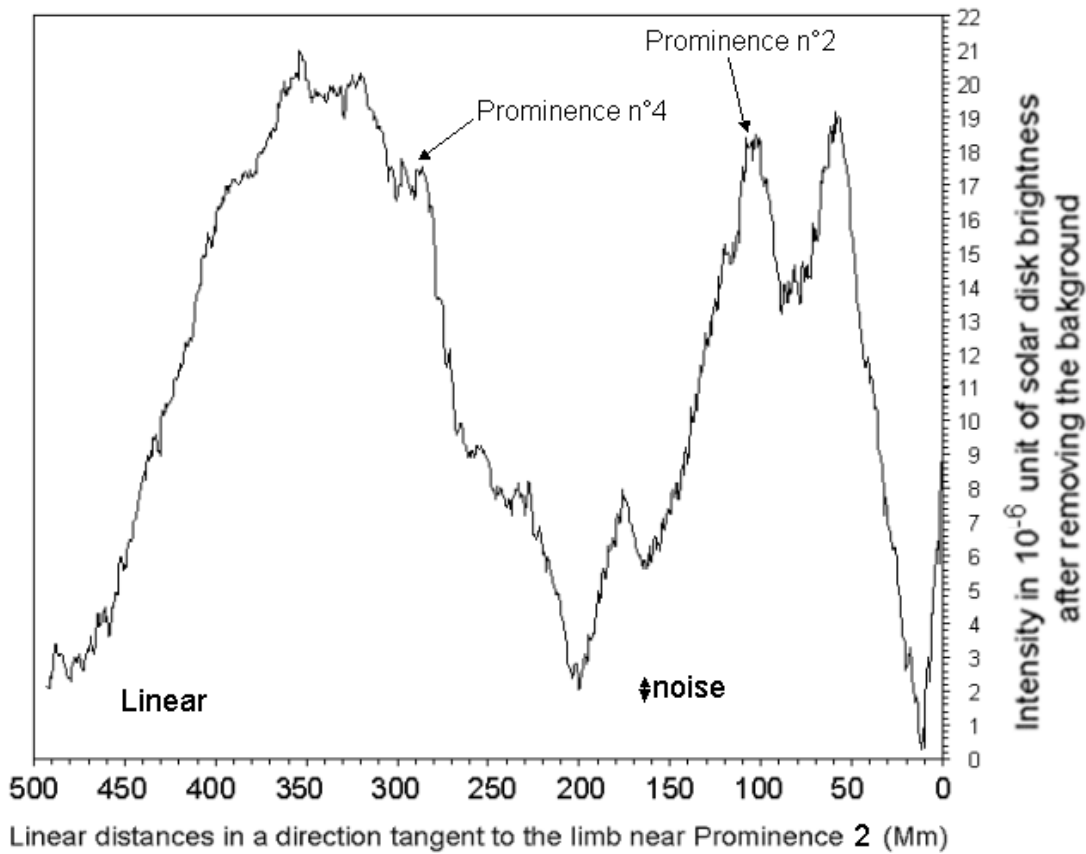

Figure 4b: Photometric cut deduced using the variation shown on Figure 4a by removing the background corresponding to the shown fitting. The relative intensity deficit around the prominence 2 is evidenced.

Figure 4a shows the result of superposing original photometric cuts taken linearly along the direction tangential to the limb (see the dashed lines in Figure 2), and not exactly along the local limb, which is obviously curved. This allows us to increase the signal-to-noise ratio in the region between the prominences and to distinguish both the true continuum of each prominence superposed on the local coronal continuum recorded along the line-of-sight and the depressed continuum outside the prominence, corresponding to the cavity. This is indeed a difficult task because of the overlapping of structures along the spectral dispersion direction when a slitless spectrum is used. Note that a selection effect similar to the effect produced by the slit of a spectrograph exists because the radial extension of all recorded emissions is drastically limited by the scale heights of each relatively cool emission line, and, to a smaller extent, the continuum of the chromosphere. The complication comes from the fact that the spectral dispersion is not exactly along the radial direction and, worse, because the dispersion direction changes with respect to the radial direction. The deviations of the observed intensity plotted in Figure 4b deduced from the smoothed fit, correspond to the continuum-intensity variations and modulations, coming from prominences seen at higher altitudes and the nearby region. Additional brightenings in He I 4713 Å and He II 4686 Å emission lines are seen as shells (Bazin *et al,* 2011) bounded on one side by the lunar profile (mountains and valleys), but they will not be considered here. Accordingly, our quantitative results on prominences alone should be considered with caution due to line-of-sight integration effect. Some evaluation of the coronal cavity effect, including the prominence -corona region seen in WL "clean" continuum near the 4686 Å helium line, has- however- been attempted (Figure 4a and 4b). The RMS noise level is estimated to be less than $1\times10^{-6}$ units of the mean solar disk intensity. The deficit is evaluated to be $4\times10^{-6}$ in units of the mean solar disk brightness in the case of prominence number 2, where the apparent coronal continuum background drastically decreases with radial distance. The deficiency is $3.5\times10^{-6}$ for prominence number 2, where the coronal background is less sharply decreasing. This represents a relative value of 7 to 12 % for prominence 2, which is almost ten times less than what we obtained using the SWAP filtergrams at 174 Å: 60 to 70% (see Figure 9).

The analysis of the deficiency needs more investigation in order to understand its relationship with the corresponding prominence. This is why we now analyze the intensity ratio of He I 4713 Å /He II 4686 Å by making a mapping, where the signal is more significant. We note in Figures 4a and 4b that in the continuum, the amplitude of the signal corresponding to the precise location of the prominences is typically of the same order as the signal of the deficit, possibly arguing in favor of the idea that the missing gas from the low parts of the cavity is filling the prominence. This is probably the first time that the true continuum of a prominence is spectroscopically compared to the surrounding coronal background, thanks to the use of this deep flash spectrum obtained after integrating almost 100 spectra. Filtergrams used in the past often contained some contribution from line emissions, which created the impression that electron densities are higher than they are in reality. Obviously, the contribution of the prominences is well recorded over the flash spectra at the location of the He lines, even in the case of a faint line such as the 4686 Å Pα line (see Figures 2 and 5).

### 2.3) Mapping of the Intensity Ratio of He I 4713 Å / He II 4686 Å

Figure 5 shows the prominences seen in the Helium lines during the eclipse totality and using the optically thick 304 Å line from SDO/AIA simultaneous filtergrams.

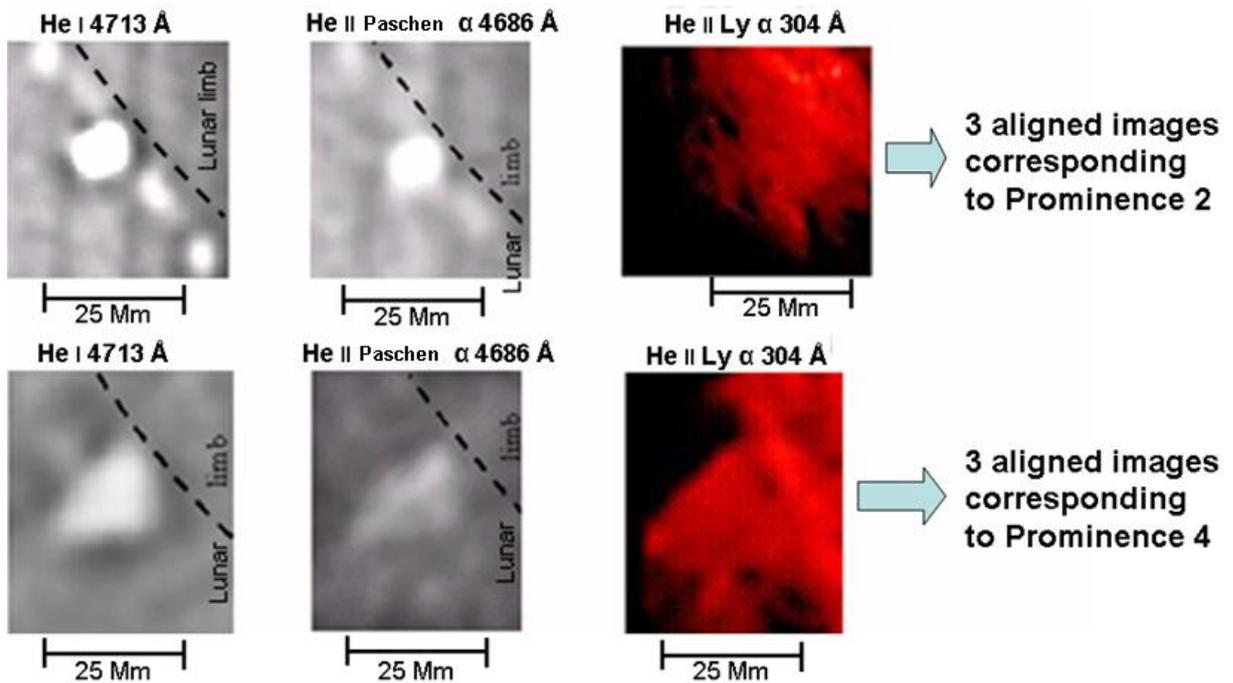

Figure 5: Images of two prominences in the 4713 Å line of He I (at left), in the Pα 4686 Å line of He II (in the middle) and in He II Ly α line at 304 Å (several order of magnitudes thicker than the Pα line).

We now consider the ratio of intensities He I 4713 Å/ He II 4686 Å which possibly reveals the temperature structure of the prominence 2 (see Figure 5) and of the surrounding region.

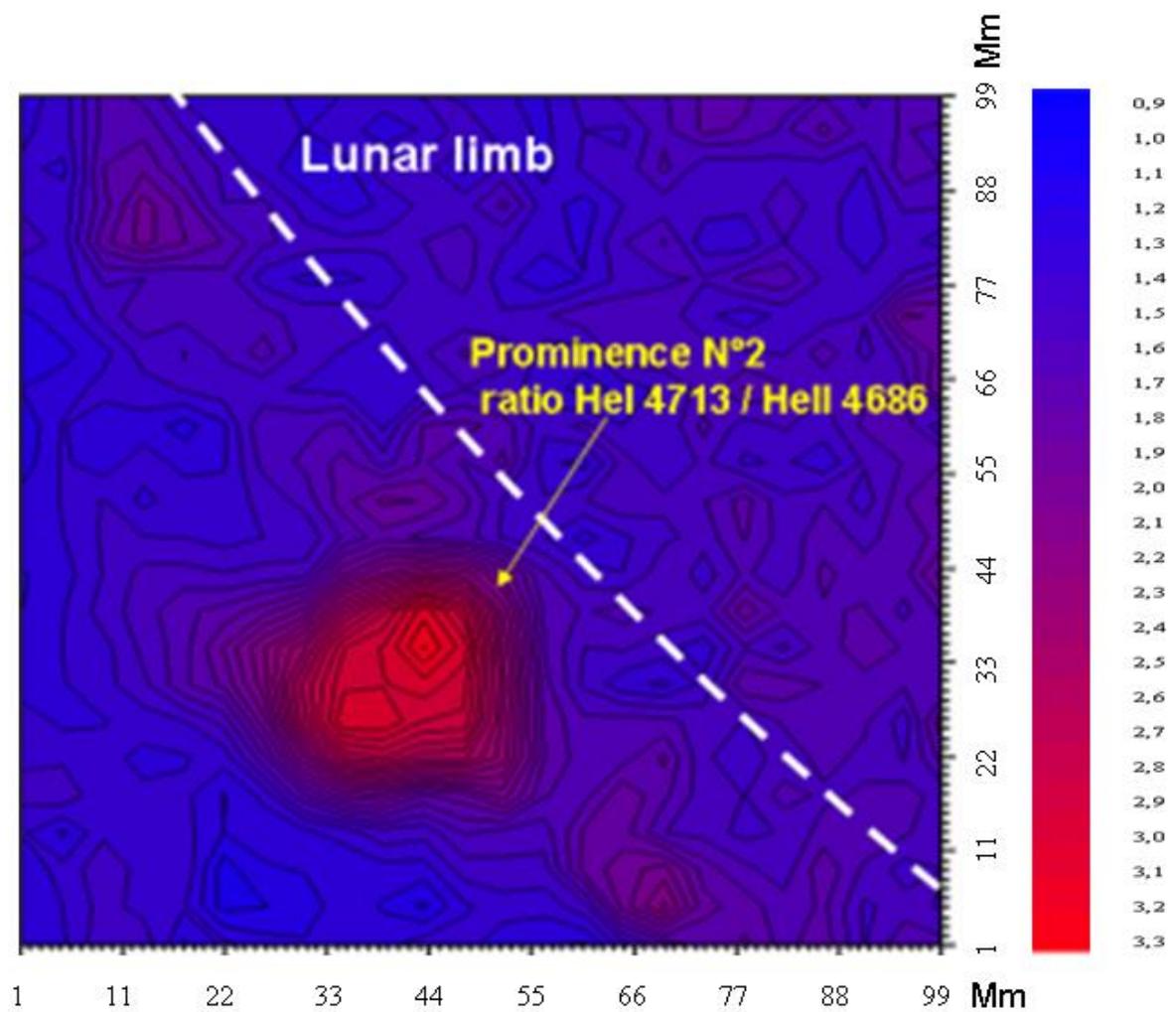

Figure 6a: 2D map showing that inside the core of prominence 2 the relative intensities of the Pα emissions of ionized helium are three times smaller than the neutral He I emissions or otherwise, that these He II Pα emissions are relatively enhanced in the outer parts of the prominence, compared to the neutral He I emissions.

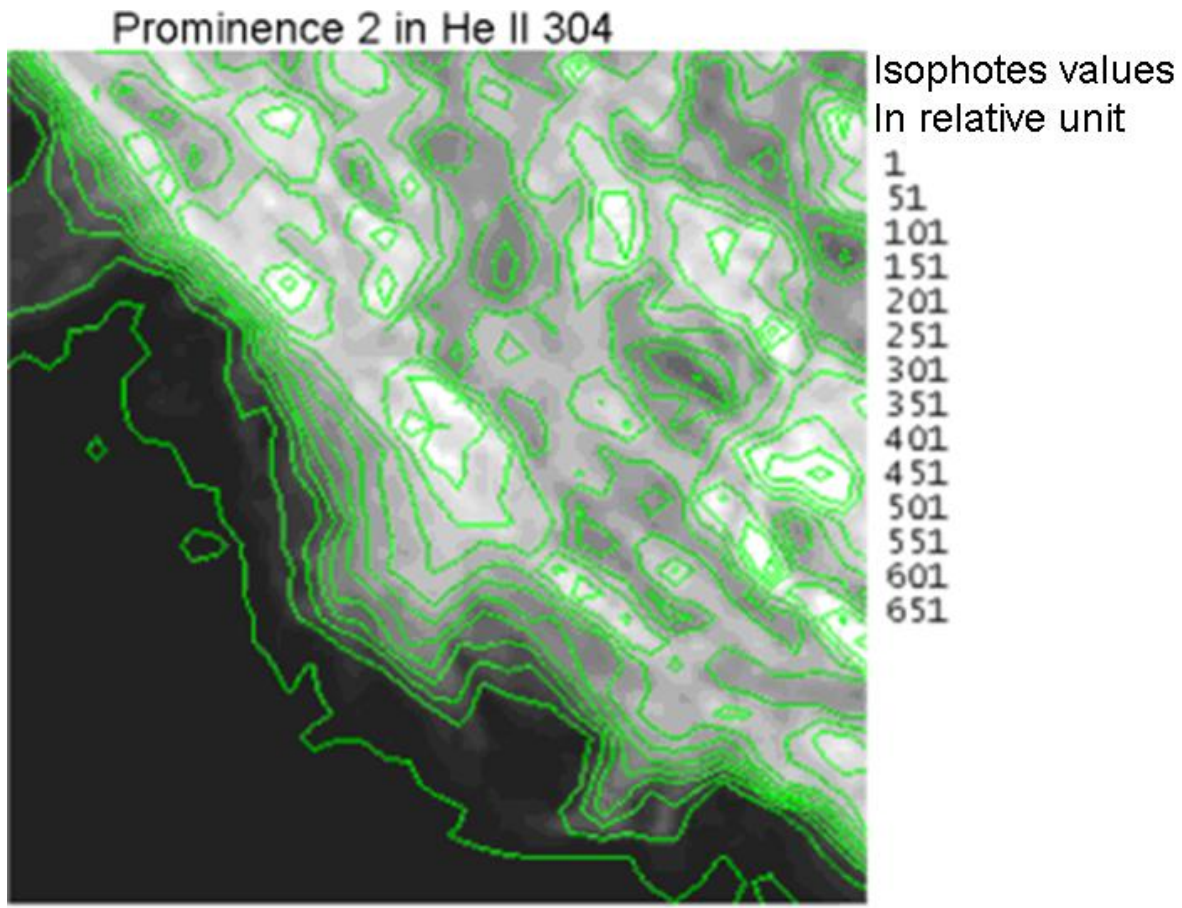

Figure 6b: Isophotes of the same prominence number 2 in the 304 Å He II line, observed at 19:19 UT with SDO/AIA.

The maximum values of the intensity ratio of He I 4713 Å/ He II 4686 Å is found to be 3.3 (Figure 6). This is in excellent agreement with the Hirayama and Irie (1984) photographic results and with what was observed in the chromosphere by the same authors, interpreted as a result of the ionization of the helium lines by UV radiation (Hirayama, 1971). The intensity ratio increases abruptly near the prominence edge, where the P -C region is located, by a factor of two over a distance of 10 Mm. This needs to be compared with the surrounding cavity, observed with PROBA2/SWAP in the Fe IX/X line at 174 Å (Figure 2), where the temperature reaches 0.6 to 1 MK (Sirk, Hurwitz, and Marchant ). Surprisingly, the SWAP image also shows, at small scale, some coronal emission near prominence 2, with the dark cavity surrounding.

## 2.4) Photometric Analysis of the Cavity Regions in WL, with SWAP and with SOHO/EIT.

An image taken in WL at the time of the 11 July 2010 eclipse totality in Hao (French Polynesia) was used. We compare the WL intensity profiles corresponding to the K-corona with SWAP 174 Å profiles, in order to analyze the cavity regions. The F-corona has been subtracted, by using the F-corona radial intensity profile from Koutchmy and Lamy (1985). The integration was performed over a 20 Mm width to get a better signal-to-noise ratio. Tangential- and radial- intensity cuts are performed around the prominences (from 110 stacked SWAP images converted to polar coordinates).

Figures 7 and 8 show the region of the cavity with the surrounding corona in polar coordinates.

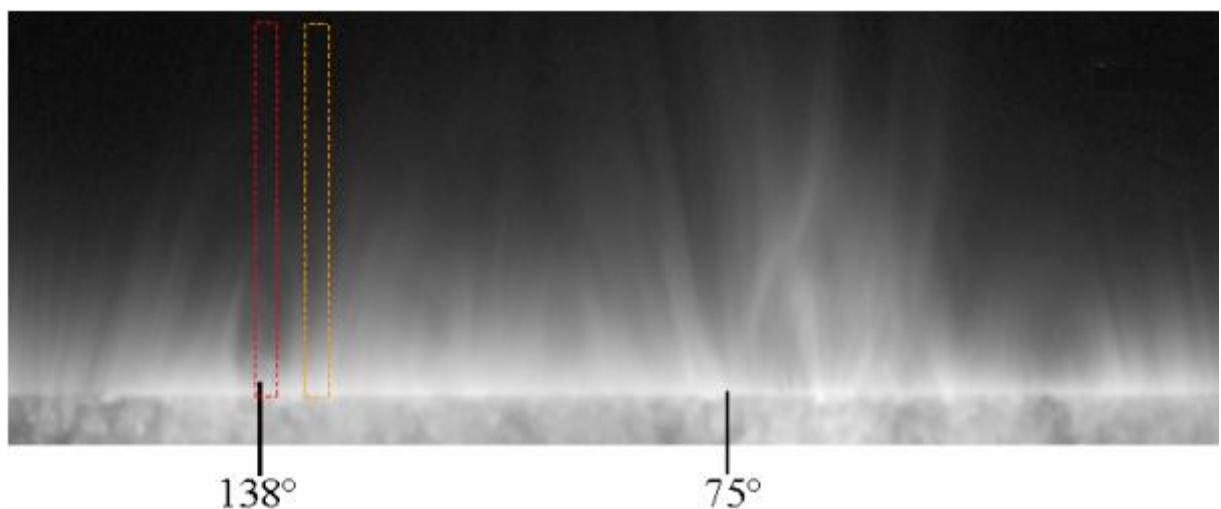

Figure 7: SWAP image taken at the time of the eclipse, in polar coordinates. The cavity of prominence 2 studied is located at 138° heliocentric coordinates. The red and orange dotted lines show where the radial cuts are taken along and outside the cavity, see Figure 10.

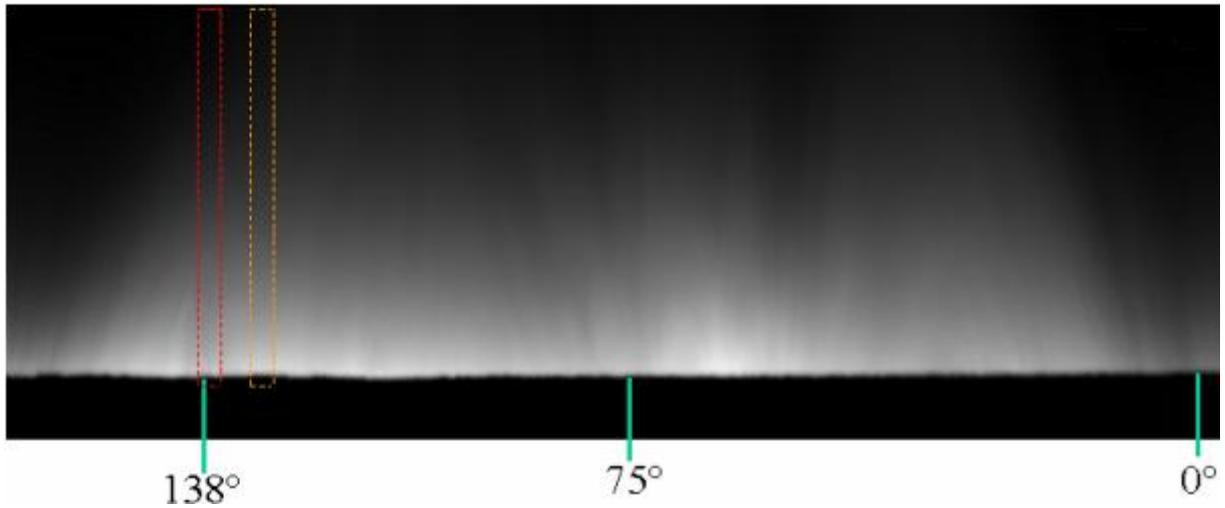

Figure 8: WL original image obtained during the 11 July 2010 total eclipse and converted to polar coordinates, showing a weaker radial gradient and the lower contrasted structures of the coronal intensities. The red and orange dotted lines indicate the position of the radial cuts, along and outside the cavity.

Figure 9 shows selected tangential cuts taken at different radial distances, between 5 Mm and 70 Mm height, from the SWAP image taken in 174 Å. It shows considerably depressed intensities of this cavity region, reaching values of order of 50 %. The deficit ratio seems to be constant for heights up to 55 Mm. Above 70 Mm, the cavity effect seems to disappear completely.

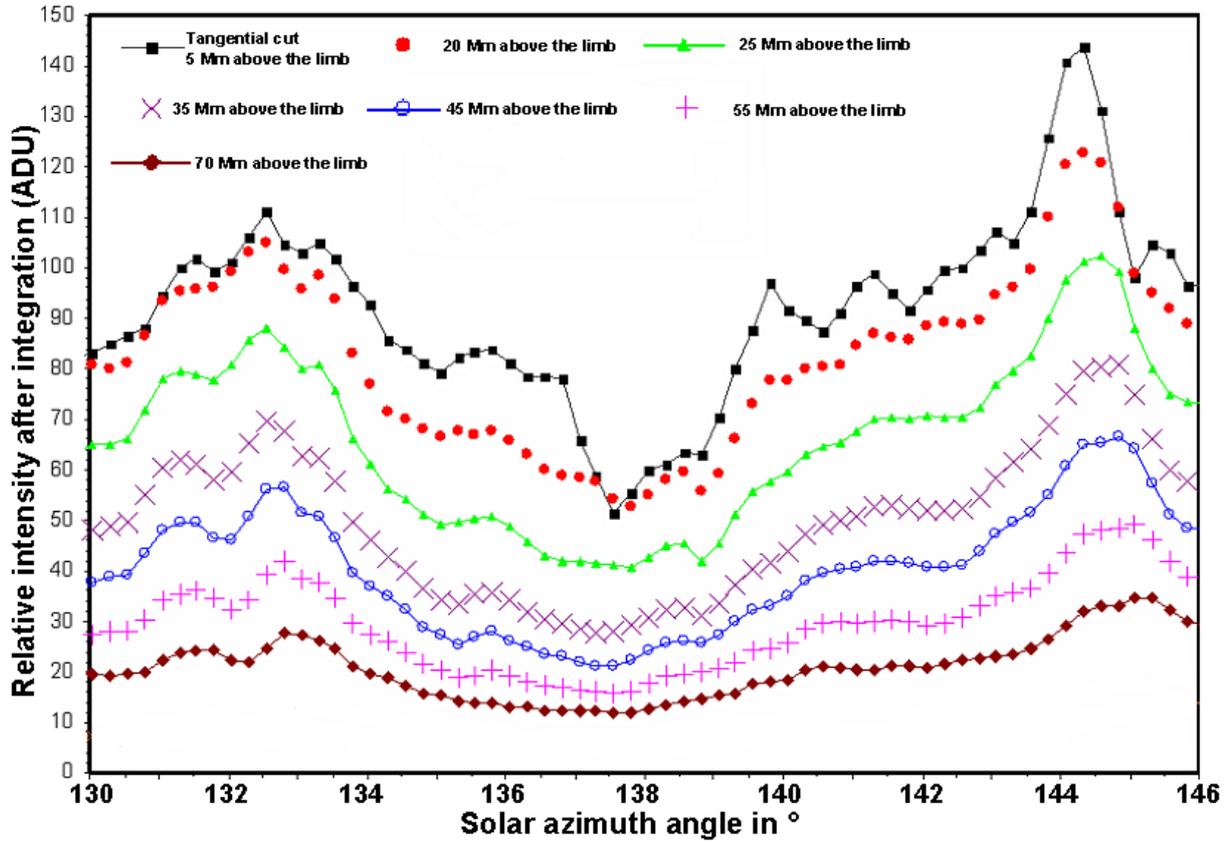

Figure 9: Tangential cuts taken approximately from 5 to 70 Mm around the SWAP cavity region showing the amplitude variations of the depression with height measured from the lunar edge.

The extension of the cavity seen with SWAP needs more analysis in order to evaluate the deficit in the Fe IX/X lines and to interpret it, using emission measurements, assuming a reasonable value of the temperature inside the cavity. In addition, line-of-sight effects need study. Figure 10, showing the radial variations, is obtained by integrating over a 20Mm width (see Figure 7). It shows the intensity profiles taken along the cavity direction and outside the cavity.

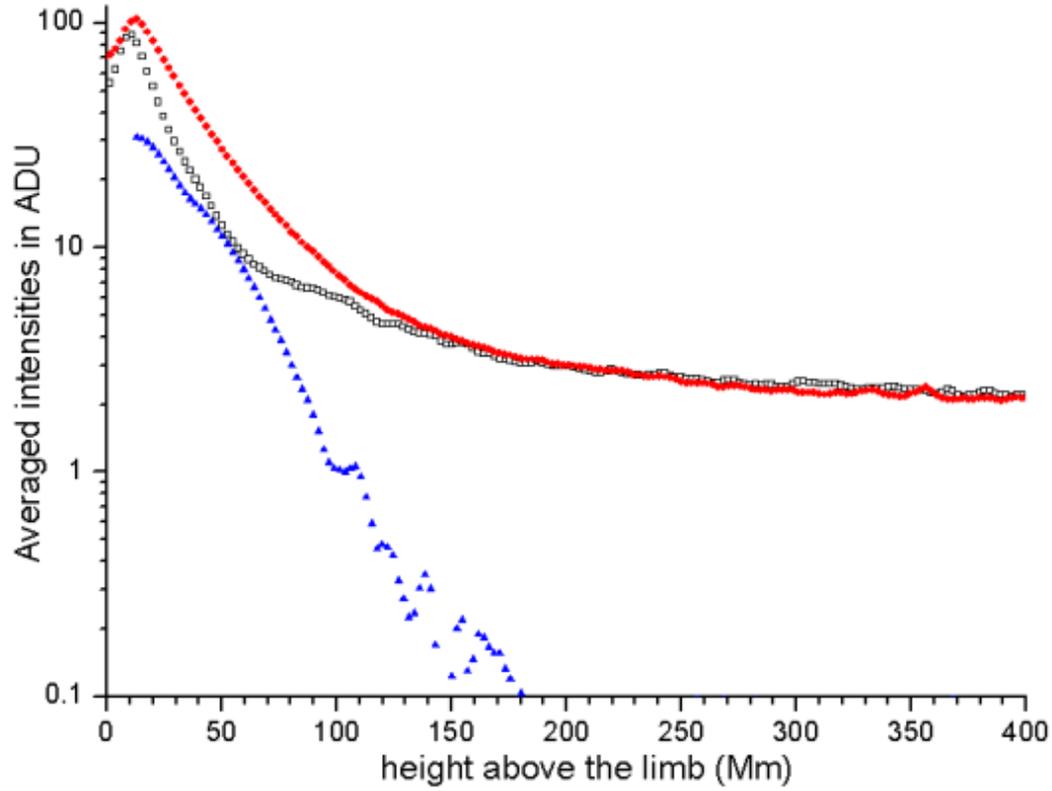

Figure 10: Radial- intensity profiles from SWAP taken in the region of the southeast cavity at 138° and difference between the radial cuts inside $I_{CAV}$, and outside the cavity $I_{OUT}$, using azimuthally a 20 Mm width of integration. The red curve corresponds to the radial cut, taken outside the cavity, the black one corresponds to the profile through the cavity. The blue curve is the difference between the radial cuts taken along and outside the cavity. The deficit shown reaches a relative maximum of about 50 % of the background value in the inner part near 60 Mm of height.

Provided the temperature inside the cavity is near the temperature $T_i$ given by the ionization balance (0.6 to 1 MK temperature), the emission in the 174 Å Fe IX/Fe X emission lines is proportional to the coronal electronic density squared [$N_e^2$] when the excitation of the line is collisionally dominated (see Allen, 1975; Thomas, 2003 and also the classical monographs by Shklovskii, 1965; Billings, 1966). In SWAP 174 Å, we find a contrast ratio $[I_{OUT} - I_{CAV}]/[I_{OUT} + I_{CAV}]$ at the height of 60 Mm of the order of 0.35 (Figures 9 and 10). The effective length of emissivity [$L_{eff}$], observed along the line-of-sight, is different for SWAP 174 Å emission and for WL intensities due to the Thomson scattering from electrons, which is only linearly proportional to $N_e$. In the case of 174 Å, intensities are proportional to $N_e^2 f(T_i)$ and accordingly, the effective length of the line of sight integration [$L_{eff}$] should be smaller (Allen, 1975) for an homogeneous corona. However, in the case of 174 Å the temperature variation along the line-of-sight is another factor that should be taken into account. In WL, the temperature shapes the corona only in the radial direction, assuming hydrostatic equilibrium, and usually this temperature is considered constant over several hydrostatic scale heights (November and Koutchmy, 1996). The radial gradient of Figure 10 is in agreement with the classical values (Shklovskii, 1965; Billings 1966) showing a decrease by a factor of ten in intensity between 1 and 1.3 solar radii. In the comparison between WL and EUV images, we incidentally found some similarity with the emission of the Fe XV line at 284 Å, as reported by Allen (1975, see his Figure 7 page 172), where a difference of the electron density deduced from WL (Thomson scattering, only) at eclipse condition and from the Fe XV emission line at 284 Å (which corresponds to rather higher temperatures) is reported. In order to explain the discrepancy, Allen (1975) introduced an irregularity factor with a value of four near the surface, but his study concerns the general corona with a spatial resolution significantly lower than the resolution of today where structures are now analyzed.

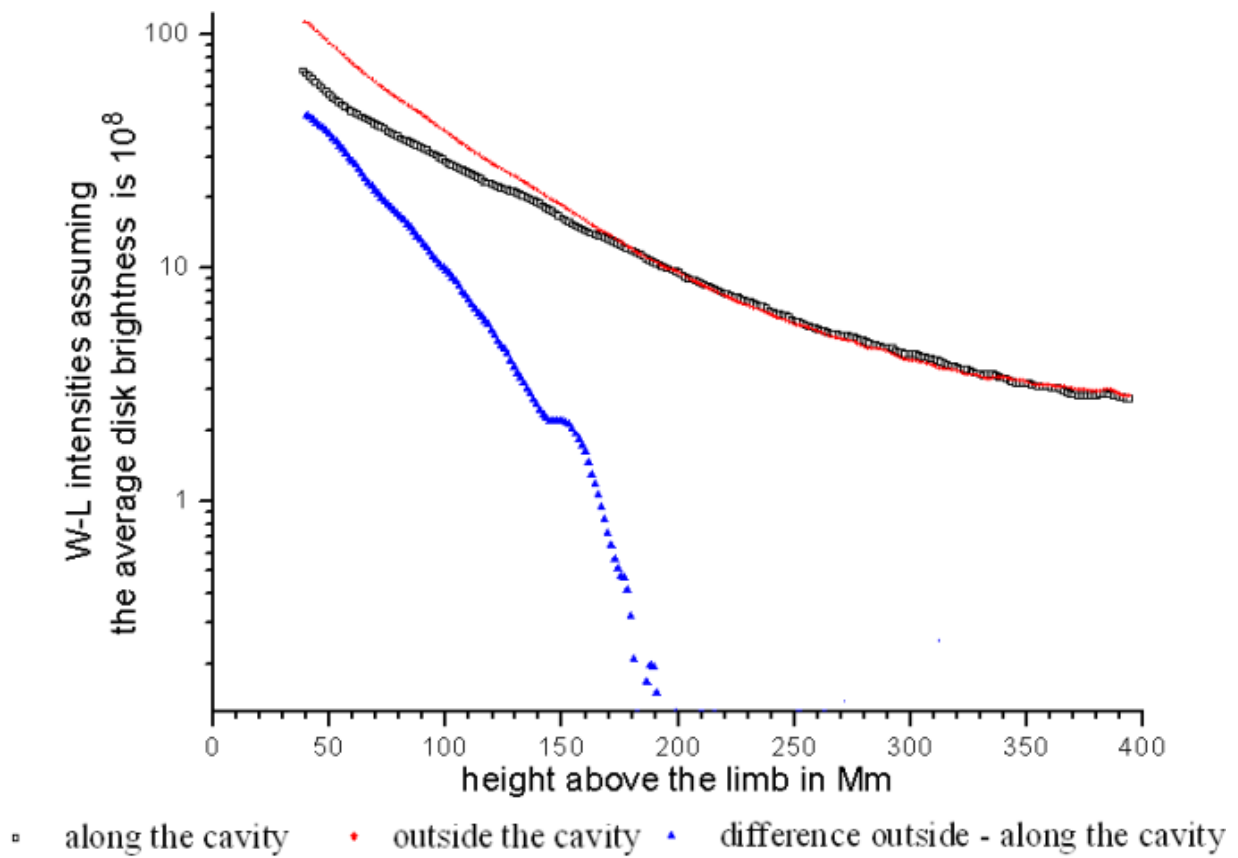

Figure 11: Radial cuts along the cavity in black, outside in red, and in blue the absolute difference along and outside the cavity, from the WL images. The contribution of the F-corona was subtracted before showing the cuts. Note that the heights of the continuum measured here is well above, in the radial direction, the height of the continuum reported from the flash spectra.

In the case of a homogeneous WL corona and for these radial distances (near 60 Mm height), November and Koutchmy (1996) computed the effective length of integration [$L_{eff}$]; it is of the order of one solar radius or 700 Mm. From the value of the intensity deficit observed in WL (Figure 11), and assuming a negligible value of the electron density inside the cavity, it is possible to evaluate the effective "extension" of the cavity along the line-of-sight, close to the horizontal direction. We also assume that the cavity is extended along the line-of-sight on the day of the eclipse. Figure 8 shows the WL azimutal intensities in solar heliocentric angles as a function of the solar radial distance. From Figure 11, in W-L at 60 Mm heights we measure a contrast value of 0.25, which is typically 1.4 times lower than what we measured in 174 Å with SWAP (Figure 10). In WL, the extension of the cavity is then evaluated to be 700x32/80 = 280 Mm along the line-of-sight, provided it is completely empty.

We also used images from the *Extreme-ultraviolet Imaging Telescope* (EIT) (Delaboudiniere *et al* 1995) of the *SOlar Heliospheric Observatory* (SOHO), from ESA and NASA. This instrument provides images of the hot corona (1.5 MK) in 195 Å Fe XII iron and in the 304 Å HeII lower temperature lines.

The field of view is 1024x1024 with 2.6 arcseconds pixels (http://umbra.nascom.nasa.gov and http://sohowww.nascom.nasa.gov).  Images in 195 Å, although noisier, allowed us to examine the cavity at higher temperatures. Figure 12 shows the 12 stacked fits images of EIT 195 Å converted to polar coordinates.

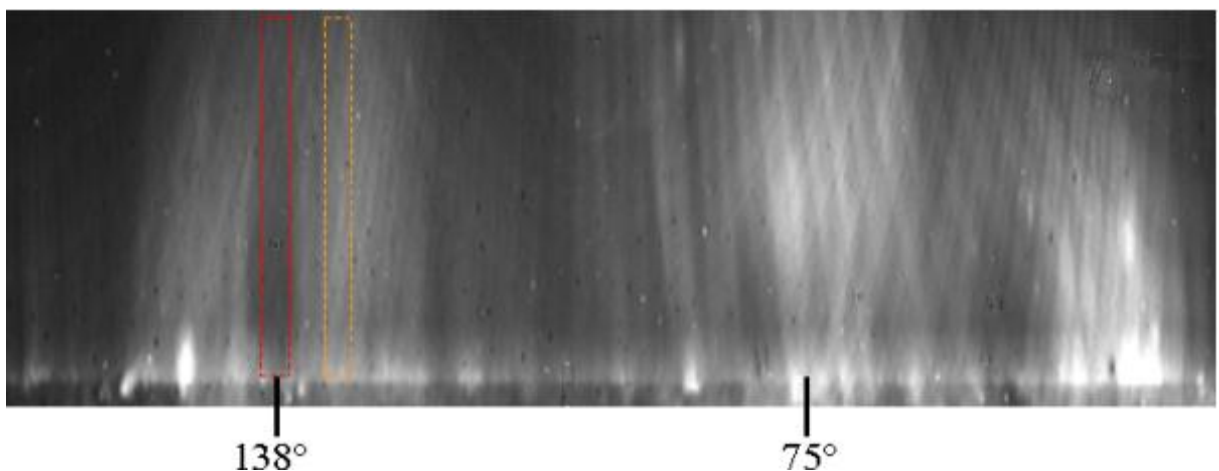

Figure 12: 12 stacked EIT 195 images to show the southeast cavity at 138° between 17:36 to 20:36 UT, and the emissivity of the surrounding corona

The deficit values at 195 Å are surprisingly found to be only about 6.6 %, without taking into account the effect of the spurious scattered light (stray-light effect). In this "hotter" emission line of about 1.5 MK the cavity contrast seems reduced; a result similar to the one recently described by Habbal *et al.* (2010, 2011) and by Pasachoff *et al.* (2011) for eclipse cavities. Moreover in Figure 12, in the radial direction, we see in 195 Å a larger scale height, or more precisely, a weaker radial gradient than seen in 174 Å, see Figure 7, provided that we trust the intensity variations given in the EIT filtergrams. This effect is also present in WL; see the radial scans shown in Figure 11. A weaker WL radial gradient corresponds to a higher hydrostatic temperature because the WL intensities reflect the values of the electron densities.

**3) Extension of the Cavity Region**

We used 20 stacked SWAP images to increase the signal-to-noise ratio, and extracted the cavity region corresponding to prominence 2, and we repeated this processing every 12 hours between 9 July and 13 July 2010, for evaluating the cavity extension, (Figure 13).

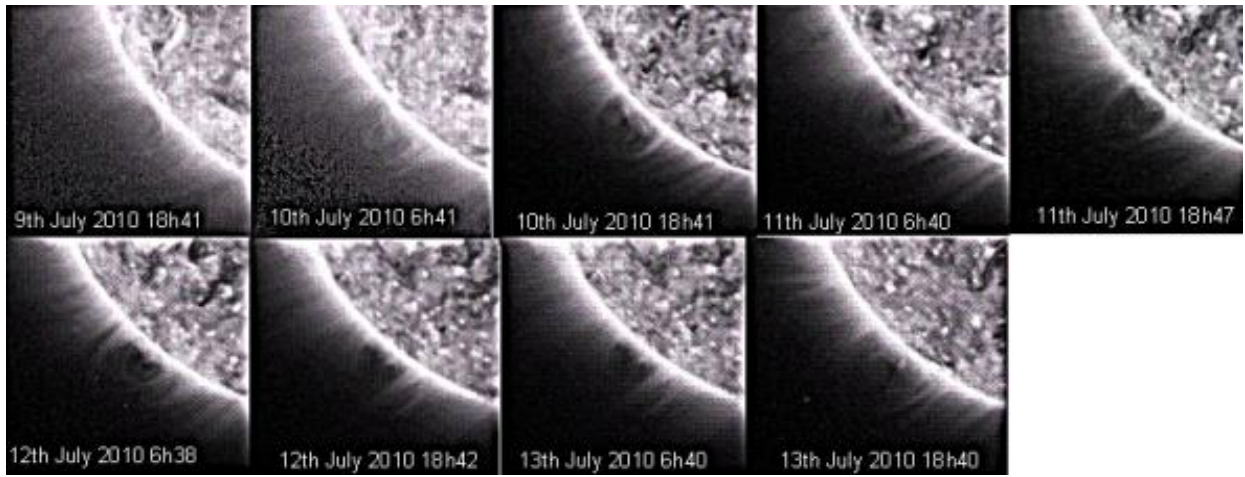

Figure 13: The cavity region of prominence 2 seen every 12 hours of time using SWAP images, from 9 July to 13 July 2010. The cavity can be better identified over the limb from 10 July near 18:00 to 13 July near 6:00 during 2.5 days. Plasma flows inside the cavity core are also present.

Although the exact geometry of the cavity is rather incertan, it seems that the limb passage is occured 1/2 day after the eclipse observation.

## 4) Discussion and Conclusion

The 2D mapping of the He I/He II relative intensity ratio shows that the maximum value (in agreement with Hirayama) is about 3.3 in the central parts of the prominence and, interestingly, that the ratio decreases abruptly by a factor of two over 10 Mm going to the outer parts, illustrating the behavior through the region where the temperature, and consequently the heating rate, are drastically increasing. A quite similar result, based on a diagnostic that uses completely different lines, was deduced by Stellmacher, G. *et al.* (2003).

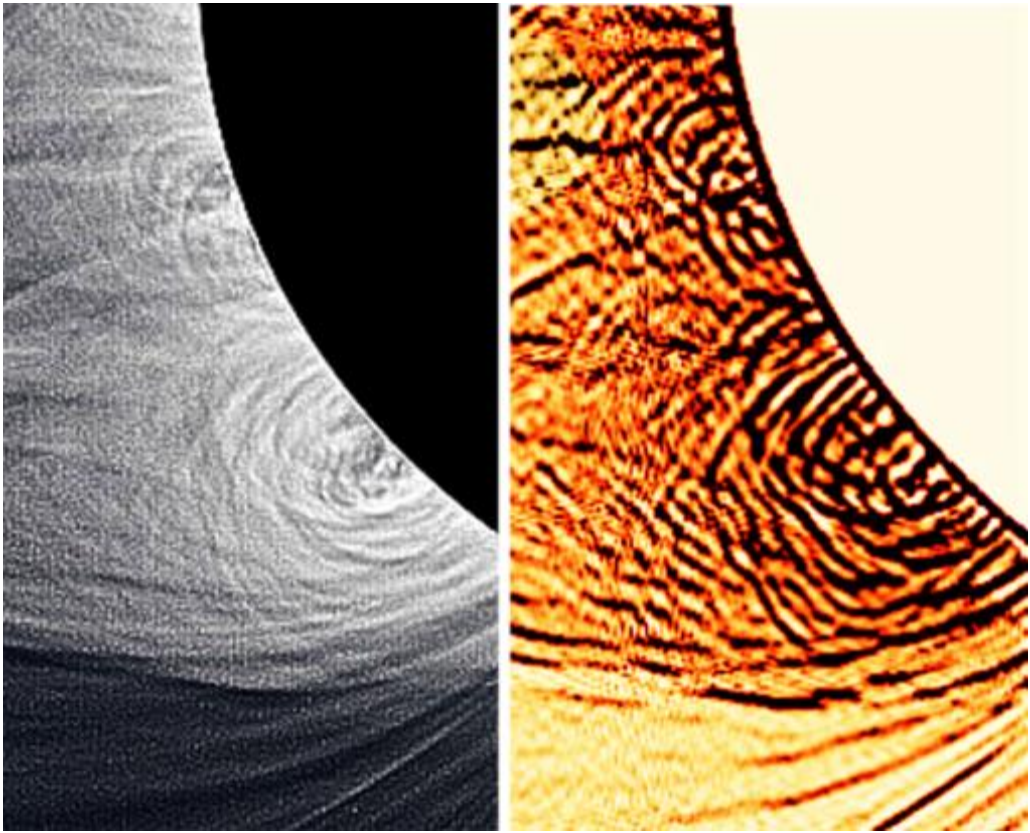

Figure 14: Extract from the highly processed WL eclipse image made for showing the contour of loops and fine-scale details after removing the larger-scale gradients (low spatial frequency filtering). At left extract from the best image provided by M. Druckmuller (Habbal *et al*. 2011) and at right, in negative, the same image after processing using the Madmax operator (Tavabi and Koutchmy 2012). Note that the processing removes the photometric properties of the picture and the cavity intensity deficit is completely washed out.

At larger scales, the photometric cuts show the typical depressed coronal intensities, in both 174 Å and WL, revealing the cavity region around the prominence with a deficit of density.

From the details of the summed flash spectra revealing the behavior of the continuum where the prominence is recorded in He lines, we obtain evidence of a rather low electron density inside the prominence, not significantly larger than that which is missing in the surrounding electron corona at the chromospheric level. This is a result of using the true continuum levels (Thomson scattering by free electrons), thanks to the use of the spectral data, and it needs to be confirmed at forthcoming eclipses.

Furthermore, the extension of the cavity along the line-of-sight was evaluated, without taking into account heterogeneity. It is of the order of 280 Mm in WL and possibly different in the SWAP 174 Å emission (0.6 to 1 MK temperature). The evaluated extension of the cavity could also depend on the temperature. Harvey (2001) reported that cavities are hot plasma inside the filament channels. An extensive discussion of the temperature effects in cavity regions was recently published by Habbal *et al* (2010, 2011). Our results tend to confirm their conclusions. In principle, the extension of the cavity-channel region can be evaluated, assuming no temporal change, thanks to rotation effects; see Figure 13. We also evaluated several movies made from AIA 171 Å images processed to remove the radial gradient in order to better see dynamical effects and, eventually, estimate the geometry of the cavity using different points of view offered by the rotation effect. They show a lot of small- and large-scale dynamical effects, in addition to showing that the cavity is indeed surrounded by many tiny irregular and moving loops that are difficult to measure. The analysis of dynamical effects is beyond the scope of this article and here we just mention the existence of the effect in the frame work of a search for the 3D structure of the cavity. The overall impression is that the cavity region, at least for the case of prominence 2, is "active"; that its extension is- however- limited, see also Figure 13, and that its geometrical extension is not very much different from what is obtained from the evaluation made using the WL diagnostics (280 Mm), without taking into account irregularities. Irregularities play an important role in case of coronal emission lines, because their $N_e^2$ dependence and, accordingly, the geometrical extension of the cavity can be longer than the effective emission length of a coronal line. In addition, temperature effects can play a role because the emission measure of each line depends on the ionization temperature of the corresponding emitting ion.

The cavity is more elongated along solar latitude than it is along the local longitudes; see Figure 13. Assuming the duration of the limb transit of the cavity is 2.5 days and taking into account its low latitude and a rotation rate of 28 days, its geometric extension is of order 240 Mm, not far from the extension deduced from the WL photometry assuming an empty cavity.

A highly processed and magnified WL image, (Figure 14), illustrates the overlapping effects along the line-of-sight, showing several arches (called loops in the case of coronal EUV filtergrams) seen in the region of prominence 2. It confirms the importance of irregularities and illustrates the difficulty of evaluating the extension in latitudes.

In conclusion, the comparison between deep SWAP and WL eclipse images is very enlightening and permits a more extended discussion of the cavity phenomenon. We plan to repeat these experiments and observations during the next total solar eclipse that will occur in Australia on 13 November 2012 and use again the SWAP data, simultaneously with the higher spatial resolution AIA images. The Sun in November 2012 should be more active than it was in July 2010, with more prominences, and we expect new results.

**Acknowledgements**

We first address many thanks to the SWAP team for the 174 Å images, which were provided by the PROBA2 Belgium consortium and we especially thank David Berghmans and Anik De Groof for their help in organizing the collaborative GI programs; the whole ROB PROBA2 team should be congratulated for succeeding in obtaining very good SWAP sequences during the 2010 total solar eclipse. SWAP is a project of the Centre Spatial de Liege and the Royal Observatory of Belgium funded by the Belgian Federal Science Policy (BELSPO). We warmly thank the SDO/AIA teams for providing EUV high-resolution images from space that we used in this article, courtesy of NASA, see SDO/AIA http://sdo.gsfs.nasa.gov and the AIA Instrument (http://aia.lmsal.com); we thank the laboratories involved in the development of this wonderful experiment, including the Smithsonian Astrophysical Observatory (Cambridge) and the LMSAL. We also thank the SOHO/EIT team for still providing good images after 14 years of operation, courtesy of both ESA and NASA, as a result of very successful collaborations. We thank Zadig Mouradian, Jean-Claude Vial, Götz Stellmacher, Eberhard Wiehr, Frédéric Auchère, and Philippe Lamy for discussions during the genesis of this article and later, and Eleni Dara and Leon Golub for reviewing it. Antoine Llebaria helped us with the IDL program to convert images into polar coordinates, M. Druckmüller provided an excellent processed eclipse image from Tatakoto, and Jean Mouette successfully took the white light images during the total solar eclipse of 11 July 2010 in French Polynesia and processed them. Observations were supported by CNES (France), in the framework of a program to complement the data collected by the PICARD space mission. Finally, we sincerely thank our referee for helping us in deeply improving the paper.

**Appendix: Schematic of the Flash Spectra Experiment used in Hao French Polynesia during the 11 July 2010 Total Eclipse.**

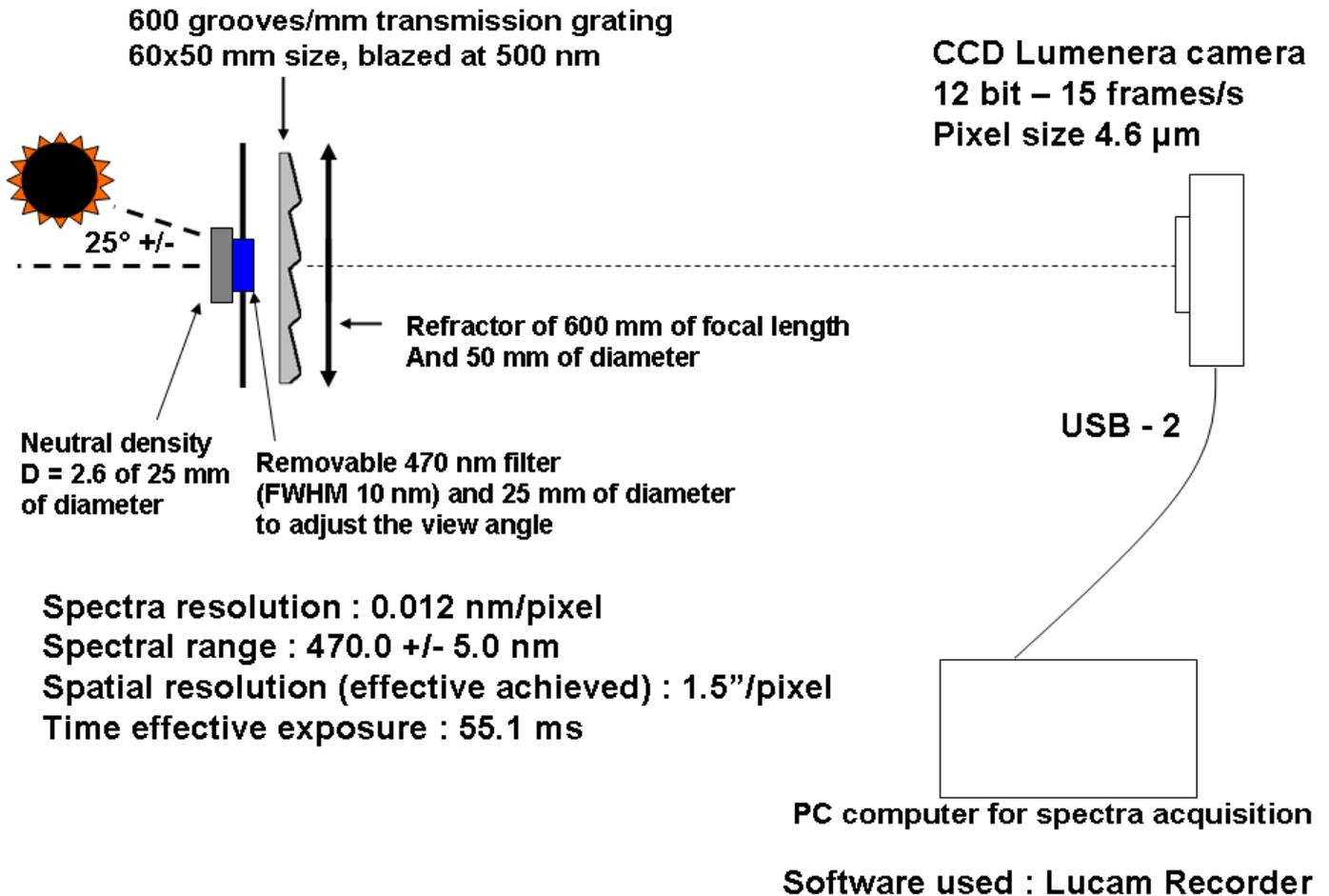

Figure 15: The setup, consisting of a transmission diffraction grating (used in first order) placed in front of the refracting achromatic lens, is fixed on an equatorial mount for guiding. The angle of approximately 25° indicates the solar spectrum deviation with the optical axis of the refractor.